\documentclass[groupedaddress,superscriptaddress,showpacs,showkeys,bibnotes,amsmath,amssymb,floatfix,reprint]{revtex4-2}
\usepackage{algorithmic}
\usepackage{graphicx}
\usepackage{dcolumn}
\usepackage{bm}
\usepackage{hyperref}
\usepackage[mathlines]{lineno}
\usepackage{mathtools}
\usepackage{blkarray, bigstrut} %
\begin{document}
\preprint{TBD}
\title{Characterization of the Firm-Firm Public Procurement Co-Bidding Network from the State of Ceará (Brazil) Municipalities}
\author{Marcos Lyra}
    \affiliation{Tribunal de Contas do Estado do Ceará, Rua Sena Madureira, 60055-080, Fortaleza, Brazil}
    \affiliation{NOVA Information Management School (NOVA IMS), Universidade Nova de Lisboa, Campus de Campolide, 1070-312 Lisboa, Portugal}
\author{Ant\'{o}nio Curado}
    \affiliation{NOVA Information Management School (NOVA IMS), Universidade Nova de Lisboa, Campus de Campolide, 1070-312 Lisboa, Portugal}
\author{Bruno Dam\'{a}sio}
    \affiliation{NOVA Information Management School (NOVA IMS), Universidade Nova de Lisboa, Campus de Campolide, 1070-312 Lisboa, Portugal}
\author{Fernando Bação}
    \affiliation{NOVA Information Management School (NOVA IMS), Universidade Nova de Lisboa, Campus de Campolide, 1070-312 Lisboa, Portugal}
\author{Fl\'{a}vio L. Pinheiro}
    \affiliation{NOVA Information Management School (NOVA IMS), Universidade Nova de Lisboa, Campus de Campolide, 1070-312 Lisboa, Portugal}
    \email{fpinheiro@novaims.unl.pt}

\date{\today}
\begin{abstract}
Fraud in public funding can have deleterious consequences for the economic, social, and political well-being of societies. Fraudulent activity associated with public procurement contracts accounts for losses of billions of euros every year. Thus, it is of utmost relevance to explore analytical frameworks that can help public authorities identify agents that are more susceptible to incur in irregular activities. Here, we use standard network science methods to study the co-biding relationships between firms that participate in public tenders issued by the $184$ municipalities of the State of Ceará (Brazil) between 2015 and 2019. We identify $22$ groups/communities of firms with similar patterns of procurement activity, defined by their geographic and activity scopes. The profiling of the communities allows us to highlight groups that are more susceptible to market manipulation and irregular activities. Our work reinforces the potential application of network analysis in policy to unfold the complex nature of relationships between market agents in a scenario of scarce data.
\end{abstract}
\maketitle

\section*{Introduction}
Despite the weight of public procurement in governmental budgets \cite{OECDStat}, it is still one of the activities that is most vulnerable to corruption \cite{murray2014procurement, OECD2015}. In that context, corruption can have many forms \cite{angulo2018eu} and occur at any point in the procurement cycle---from the pre-tendering to the tendering and the past-award phases---making it difficult to detect and measure \cite{Mufutau2016Detection,Rustiarini2019why,Whiteman2019fraud}. In Brazil alone, corruption in procurement contracts can represent an additional $20\%$ to $30\%$ of the expected price, which represents losses of around 200 billion Reais annually \cite{zeferino2020corrupccao}. Likewise, in Europe, it is estimated that losses are of around 5B Euros annually \cite{hafner2016cost}. Naturally, these losses undermine the ability of governments and public authorities to push-forward essential investments in health, education, infrastructure, security, housing, and social services \cite{soreide2002corruption, Beittel2019Combating}. Unsurprisingly, there is a considerable effort to develop analytical solutions to understand and mitigate the effects of corruption in the public procurement process \cite{fazekas2018inovations}. Recently, the increasing availability of open data concerning public administration activities \cite{curado2020scaling} has renewed the scientific community’s efforts on uncovering the hidden connections between participating agents and how their relationships can link to fraudulent activities \cite{Herrera2019network, kertesz2021complexity}.

One of the most challenging aspects of identifying corruption in the context of public procurement contracts is related to the lack of labeled data. Hence, it is often impossible to know which instances correspond to corruption \cite{wachs2019network}. In that sense, past works have approached this problem from an unsupervised learning perspective, meaning that they look to extract from the data more information about the relationships between the involved parties and, thus, flag groups of agents with patterns associated with a high risk of corruption. In that sense, a fundamental principle in public procurement is that of transparency in bidding \cite{adjei2019role, angulo2018eu, nowrousian2019combatting, spagnolo2012reputation}. In that sense, competition leads to greater efficiency for the public sector. As such, firms that developed the necessary relationships to achieve leverage to manipulating a tender process at a high risk of corrupting the procurement process \cite{hanak2018analysis}.

\begin{figure*}[!t]
	\centering
	\includegraphics[width=.95\textwidth]{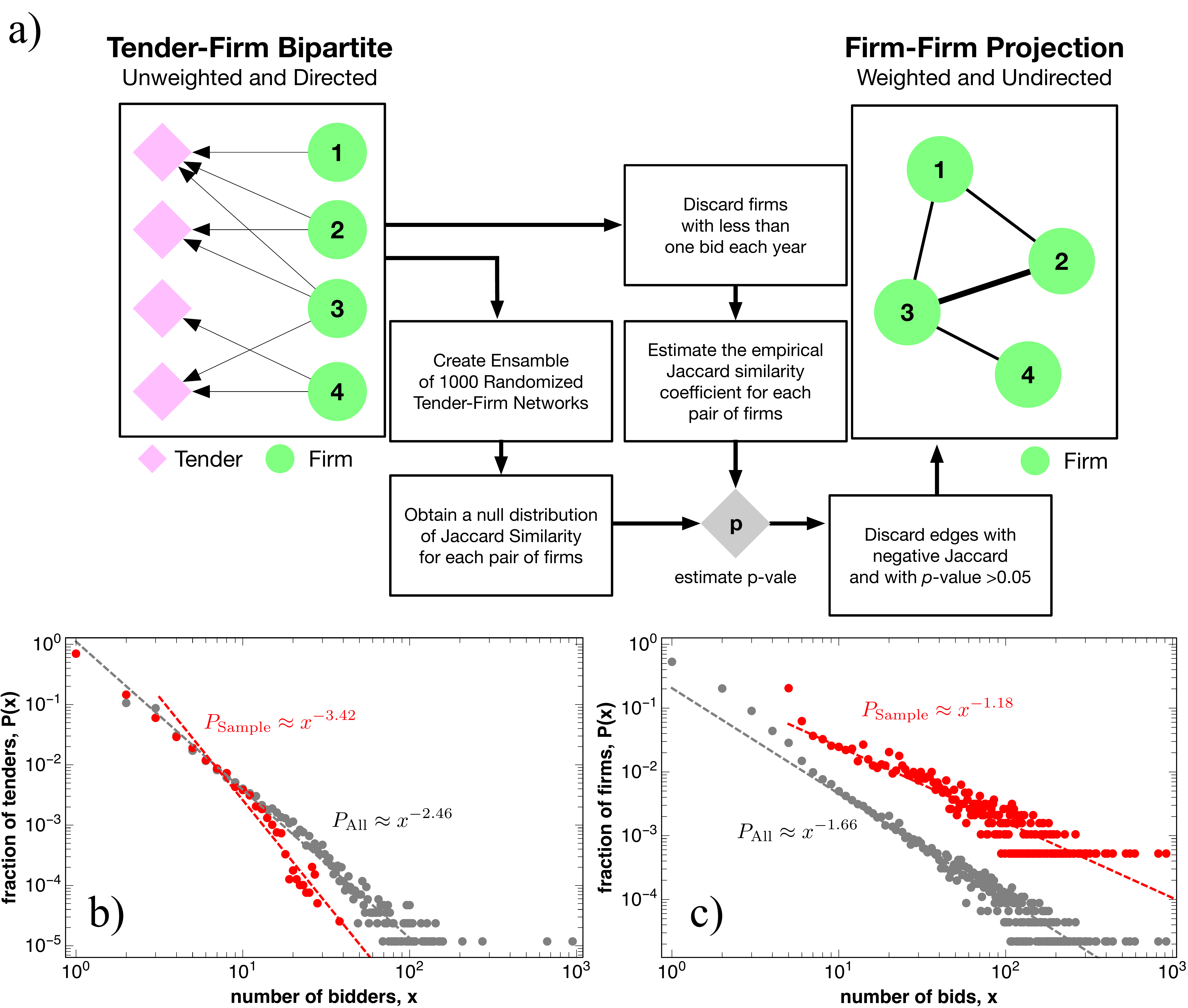}
	\caption{ Panel a), graphical representation of the process employed to infer the Firm-Firm co-bidding network. Panel b), comparison between the frequency of bidders per tender in the original data set (gray) and in the working data set (red) after filters have been applied. Panel c), comparison between the frequency of bids per firm in the original data set (gray) and in the working data set (red) after filters have been applied. In panel b) and c) dashed line represents the OLS regression lines, the domain of the line indicates the domain used for fitting the curve.
	\label{figure1}
	}
\end{figure*}

An open issue remains, can the communities identified from firms co-biding patterns allow us to highlight groups of firms that are more susceptible to collusion and market manipulation? The use of network analysis for the study of corruption is not new \cite{lauchs2011corrupt, chang2018understanding, grassi2019betweenness}. In the context of public procurement, past studies can be divided into two main groups: 1) works that explore bipartite relationships between public bodies and firms \cite{fazekas2016corruption, wachs2019social}; and 2) studies that explore firm-firm co-biding relationships in public tenders \cite{Toth2014toolkit, reeves2017bid, morselli2018net, wachs2020corruption}. Both approaches have their merits, and each is more suitable to identify different mechanics underlying the manipulation of the procurement process. For instance, bipartite relationships are suitable to identify fraud that stems from bribes and influence ties; while firm-firm relationships are more suited to identify cartels and collusion. Despite these, the use of network analysis to study the relationship between firms in procurement bids is a relatively new venture \cite{reeves2017bid}, and more evidence is required in order to have a clear picture of the universality of existing patterns and mechanics across cultural and socio-economic contexts.

Here, we use methods from network science and complexity sciences to map and characterize the co-bidding network \cite{wachs2019network, Piccolo2017design, Ramalho2020detection, reeves2017bid} between firms that participated in public tenders issued by the $184$ municipalities of the state of Ceará (Brazil). In that sense, we provide a characterization of the relationships between competing firms and identify the major communities of firms that often compete for tenders with similar scope. Moreover, we argue that some of such communities have characteristics that place them at a higher risk of market manipulation and irregular activities often associated with corruption. 

\section*{\label{data}Data}
We use data from \textit{Tribunal de Contas do Estado Ceará} (Brazil) covering public tenders issued by the 184 municipalities of the State of Ceará between 2015 to 2019. Each observation informs about the bid of a firm to a tender and whether the bid was one of the winning bids. It also includes information about the municipality that issued the tender, and whether a firm won a contract. Hence, the data is naturally represented through a bipartite nature \cite{fierascu2017Networked}, which connects firms to tenders (see Figure~\ref{figure1}a). The data set contains $196,608$ observations that account for the bids of $45,502$ firms to $84,835$ tenders.

Information about the firms and tenders is anonymized, and bidding values are not available. Moreover, the data set does not contain information about which contracts/firms have been investigated in the past for irregularities.

\section*{\label{data}Network Inference}
Since we are interested in studying the relationships between firms we focus on the Firm-Firm projection. We estimate the projection from the co-bidding patterns of firms \cite{Piccolo2017design} using the Jaccard similarity coefficient \cite{veech2013probabilistic,mainali2017statistical,chung2019jaccard,wachs2019network}. Figures~\ref{figure1}a shows a graphical illustration of the structure of the data and depicts the steps conducted in order to infer the Firm-Firm network from the original Tender-Firm bipartite structure. 

In order to infer the Firm-Firm co-bidding network, we start by discarding all firms that did not bid at least once during each year of analysis. By doing so, we are able to extract the core of active firms, while removing firms with sporadic activity. Figure~\ref{figure1}b and \ref{figure1}c compare the original ($P_\text{ALL}$) with the filtered data set ($P_\text{Sample}$). In particular, showing that filtering tends to remove excess participants from tenders, while it does not affect the distribution of the number of bids done by each firm. Likewise, we refer to firms present in the firm-firm co-biding network as Established firms. The final working data set includes $1,906$ firms, which account for $72,078$ bids to $39,523$ tenders.

\begin{figure*}[!t]
	\centering
	\includegraphics[width=0.95\textwidth]{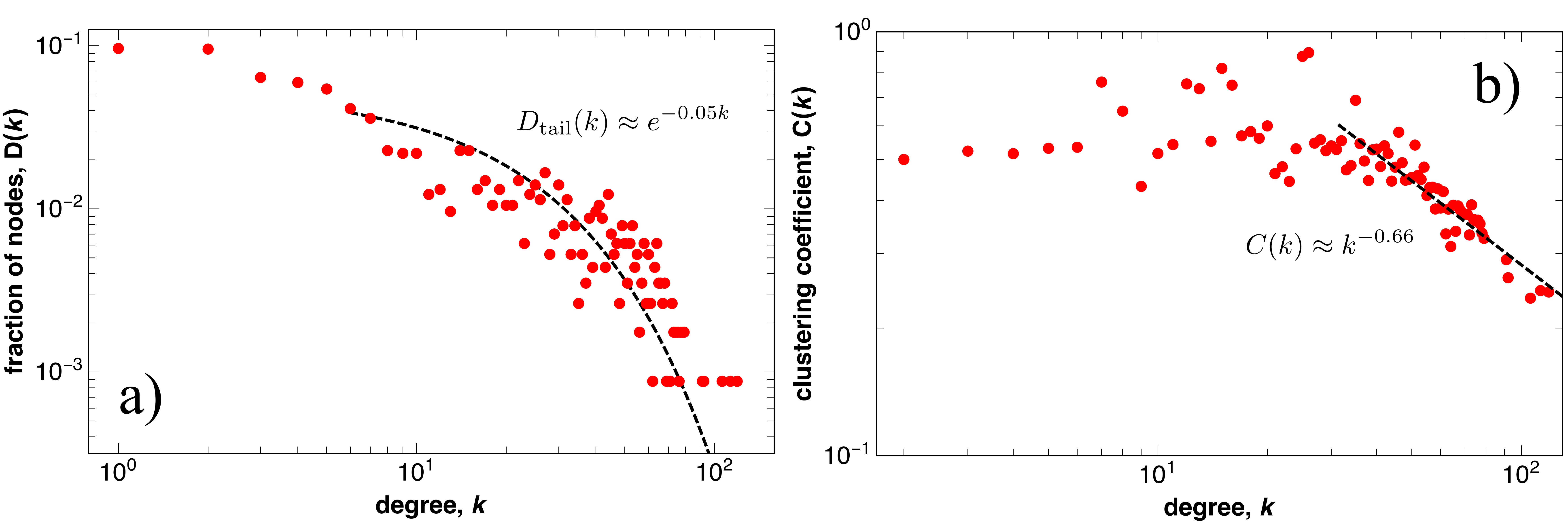}
	\caption{Panel a) shows the degree distribution, $D(k)$. The dashed black line represents the best exponential distribution fitted to the tail ($k>5$) of the empirical distribution. Panel b) shows the average clustering coefficient average per degree, $C(k)$. The dashed black line shows the best linear fit.
	Results have been estimated from the entire graph.
	\label{Figure2}
	}
\end{figure*}

Hence, next, we compute the centered Jaccard coefficient \cite{chung2019jaccard} is between each pair of firms, which can be computed as:
\begin{equation}
    J^c_{ij}=\frac{\sum_kb_{it}b_{jt}}{\sum_k(b_{it}+b_{jt}-b_{it}b_{jt})} - \frac{p_i p_j}{p_i+ p_j - p_i p_j},
    \label{eq1}
\end{equation}
where $b_{ik}$ is one if if firm $i$ made a bid to tender $t$, being zero otherwise; and $p_i$ is the fraction of tenders in which firm $i$ participated ($p_i = \sum_k b_{it}$. The second term in Equation~\ref{eq1} provides the expected number of observations when the bids from both firms are independent and identically distributed through a Bernoulli process \cite{chung2019jaccard}. Hence, the centered Jaccard coefficient allows us to distinguish between positive and negative associations between firms.

Finally, we estimate the significance of the computed $J^c_{ij}$ (i.e., to test the hypothesis that $J^c_{ij} > 0$). To that end, we bootstrap a null distribution ($\hat{J}_{ij}$) of centered Jaccard coefficient for each by generating an ensemble of $1000$ randomizations of the initial bipartite network. Data was randomized in order to ensure that the number of bids observed per firm and per year remained constant while keeping constant the number of firms bidding to each tender. Then, using statistical inference methods \cite{gotelli2000null}, we estimate the p-value associated with $J^c_{ij}$ by calculating the upper tail probability of obtaining a value equal or greater than $J^c_{ij}$ from the cumulative frequency of the null-distribution $\hat{J}_{ij}$. We discard links with $p$-value $>0.05$.

\begin{figure*}[!t]
	\centering
	\includegraphics[width=.95\textwidth]{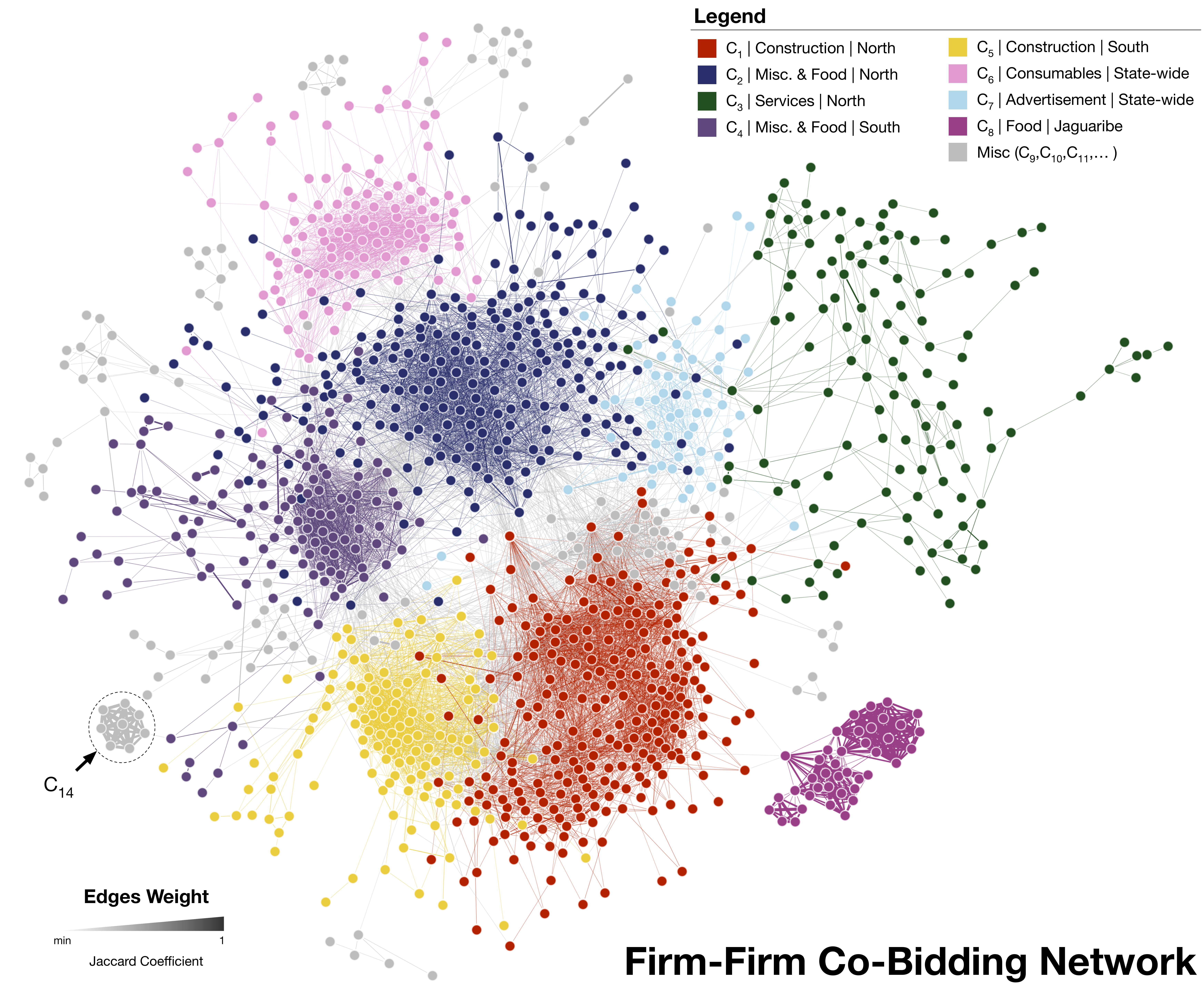}
	\caption{Graphical representation of the Firm-Firm network, which relates firms with similar bidding pattern. In order to build the network we consider only the most active firms, and edges with a significant Jaccard similarity index. Represented is the giant component with some relevant disconnected components. Nodes in the giant component are colored according to one of the eight major communities (out of $22$) identified using the Louvain algorithm \cite{blondel2008fast}. The presented partition of the network achieves a modularity of $0.61$.
	\label{Figure3}
	}
\end{figure*}

The resulting firm-firm co-bidding network contains $1,529$ nodes and $12,892$ edges. Relationships are treated as undirected and unweighted, identifying firms that have a similar pattern of bidding. The network exhibits an average degree of $16.86$, with a cluster coefficient of $0.52$ \cite{newman2020improved}, and $56$ connected components. Figure~\ref{Figure2} shows the Degree Distribution (panel \ref{Figure2}a) decays exponentially with the degree. Meaning that the underlying mechanics of co-bidding can be approximated with a random attachment process \cite{albert2002statistical}. However, the average clustering coefficient shows an inverse relationship with the degree (panel \ref{Figure2}b), suggesting the existence of some level of hierarchy in the structure of the network. It is noteworthy to mention that the largest connected component contains $1,141$ nodes, $10,630$ edges, and a clustering coefficient of $0.43$.

\section*{\label{results}Results and Discussion}
Figure~\ref{Figure3} presents the giant component of the firm-firm co-bidding network. In particular, we highlight the eight largest communities (nodes are colored accordingly). Using the \textit{Louvain} algorithm \cite{blondel2008fast} we identified $22$ communities with a modularity of $0.66$. We refer to these communities as $C_1$, $C_2$,...,$C_{22}$, whose index is ordered in descending order to the number of firms in the communities. The high modularity of the network is, however, unsurprisingly and can be explained by the fact that the network represents firms in different markets characterized by the different nature of contracts (e.g., works, services, etc) in different regions. As indicated in Figure~\ref{Figure3}, the largest communities divide the network into two major groups of firms that operate mostly in the northern (Red, Blue, and Green) and south (Purple and Yellow) regions of the state of Ceará, but also on contracts that deliver Food services (Blue and Purple) or construction works (Red and Yellow). Interestingly, the remaining communities highlighted identify firms that compete at state-wide level (Pink and light Blue) and one particular instance of a community (Violet) operating in a small region (Jaguaribe) and a single contract type (Food services). Hence, the network highlights competing markets and provides a characterization that is interpretable.

As discussed above, communities in Firm-Firm network structures can be used to identify market segments. In limiting scenarios, cases in which firms form "echo chambers" or highly dense communities, it also allows us to flag groups of firms that present a high risk of collusion and procurement manipulation. In other words, corruption. As such, next, we explore the use of network science as an approach to classify the communities of firms that might be of interest to investigate deeper by the authorities of interest. 

\begin{figure*}[!t]
	\centering
	\includegraphics[width=0.95\textwidth]{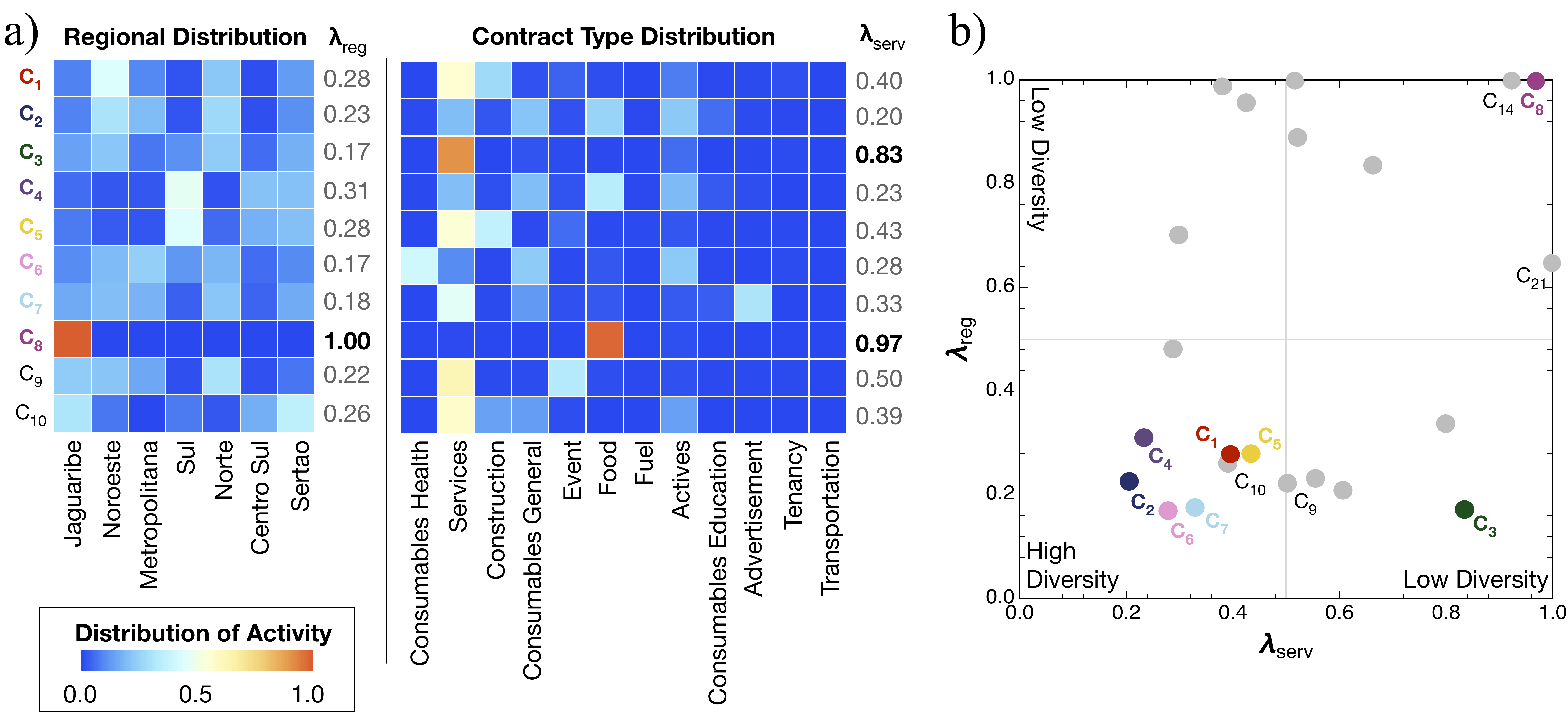}
	\caption{Characterization of the ten largest communities by the diversity of bids done by region and type. Panel a) shows the distribution of bids within each community by geographical region (left) and contract/tender type (right). For each community we compute the Simpson's diversity index ($\lambda_\text{reg}$ and $\lambda_\text{serv}$). Panel b) compares communities by their diversity of contracts in terms of regional span and type. Note that in panel a) we only show results for the ten largest communities, which are representative of the results. Communities not identified by a color code in Figure~\ref{Figure2} are shown in gray in panel b), in the particular community $C_{14}$ corresponds to the gray clique easily identifiable in the bottom left of the network in Figure~\ref{Figure3}.
	\label{Figure4}
	}
\end{figure*}

\subsection*{Activities Diversity}
We start by looking at the regional diversity on which firms performed their activities (\textit{e.g.}, bid on tenders), and also the diversity of the type of contracts that they bid to. While a firm with low diversity in both regional reach and contract-type can simply indicate a firm that is narrow in both scope and domain, the existence of clear groups (i.e., a community of firms) that share such indicator can highlight more a troublesome scenario. In particular, it can indicate the conditions for firms to coordinate and cooperate to control a specific market in a specific regional context, and should be investigated with further care. To that end, we start by estimating the Simpson's diversity index\footnote{In some fields the Simpson's index is also known as the Herfindahl index.} for each community, which can be computed as
\begin{equation}
    \lambda^{C_i}_\text{cat} = \sum_{t \in \gamma} (p_{C_i}^t)^2
\end{equation}
where $p_{C_i}^t$ corresponds to the fraction of bids done in a procurement contract of type $\gamma:$\{Consumables Health, Services, Construction, Events, Food, Fuel,...\} or regions type $\gamma:$\{Metropolinana, Norte, Sul, Noroeste,...\} by the firms in community $C_i$, $\forall i \in \{1,2,...,21,22\}$. The quantity $p_{C_i}^t$ is normalized per community, so that $\sum_{t} p_{C_i}^t = 1.0$. We estimate $\lambda^{C_i}_\text{cat}$ independently for each community ($C_i$), and for contracts according to the region that issue the tender and the tender contract type (e.g., services, food, tenancy, constraction, etc). Our choice of the Simpson's index over other alternatives (e.g., entropy) is due to its straightforward interpretation: the probability that two bids from a community are in the same category (e.g., region or contract type).

Figure~\ref{Figure4}a illustrates the empirical distributions ($p_{C_i}^t$) of procurement activity for the ten largest communities. We show the results for both the Regional distribution of activities and by  Contract Type. Blue colors denote a low relative frequency of bids, while red identifies a high frequency. These indicators allow us to infer the degree of specialization and agglomeration of a community. In particular, we find that Community 8 ($C_8$) activities are agglomerated in a single region (Jaguaribe) and firms specialize in one type of contract (Food). The same conclusion can also be inferred from the high levels of $\lambda^{C_8}_\text{cat}$, which means that Community 8 has low diversity of activity distribution. Figure~\ref{Figure4}b compares all the $22$ communities in terms of the two diversity indicators defined above. We find a clustering of communities in the bottom left quadrant--- a low level of agglomeration and specialization---that we associate with healthy markets composed of firms that, on average, have a diversified portfolio of activities and regional distribution. In contrast, in the top right quadrant, we find communities that rely on procurement contracts of a single type and agglomerated in a small number of regions.

The combination of these two diversity indicators, at the community level, provides a powerful feature to identify groups of firms that can dominate over a niche market or, in the worst case, develop undesirable leverage, as a group, in negotiating procurement contracts. Hence, lowering the desirable efficiency that public procurement aims at achieving in the tendering process. However, it is important to stress that these metrics are just indicative of potential problems, and thus the true nature of the activities of the firms in each community should be carefully investigated by the corresponding local authorities.

\subsection*{Bidding Coordination}
\begin{figure*}[!t]
	\centering
	\includegraphics[width=0.95\textwidth]{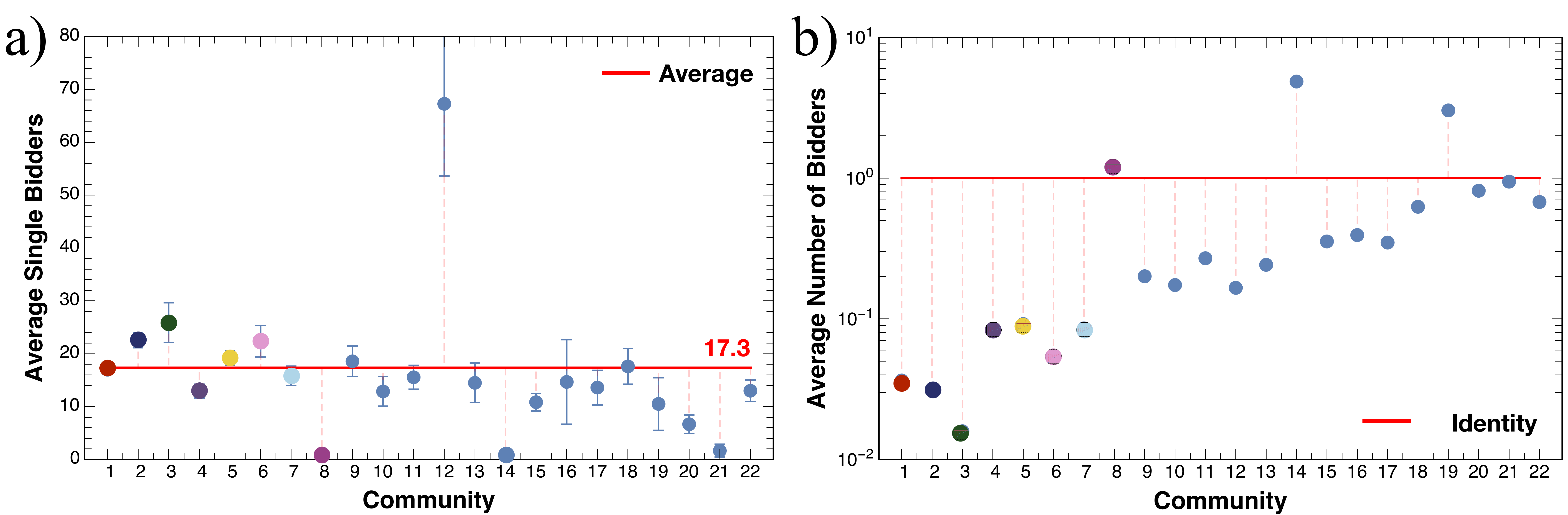}
	\caption{Characterization by bidding activity. Panel a) shows the average number of single bids per community (i.e., the average of the number of times a firm in a community participated in a tender as the single bidder). We compare the values of each company with the average of the entire population of firms (horizontal red line). Panel b) shows the average number of bidders in tenders that firms within a community typically participates. We normalize the value obtained for each community by the number of firms in that community. The horizontal red line shows the threshold that marks the size of the community
	\label{Figure5}
	}
\end{figure*}
To further investigate the risk/susceptibility of market manipulation by firms, we next look at the propensity that each community has in participating in "single bidder" contracts. Another pattern often associated with corruption and loss of efficiency. Hence, what is the susceptibility of each community to such practice? To answer that question, we start by investigating the average number of times, per community, that a firm is the single bidder of a tender. Figure~\ref{Figure5}a shows the results for all $22$ communities in the largest component of the Firm-Firm network. Traditionally a higher level of single bidders would be the main indicator of market manipulation. However, at the community level, both very low and very high levels are indicative of unusual activity. In particular, if we set as a baseline number of single bids by a typical firm. The reason being that low levels indicate the risk of coordination (e.g., firms participating coherently in the same contracts) while high levels can sign the prevalence of less competitive markets. Overall, of the largest 10 communities, only Community 8 exhibits low levels of single bidders, a pattern extended to Communities 14 and 21 as well.. In contrast, we see that community 12 strongly deviates from the average with average value of single bidding that is roughly four times larger than the average.

In addition, it is important to look at the average number of bidders per tender in order to assess the potential existence of coherent behavior, that is, coordination between the firms in a community. To that end, for each community, we estimate the average number of bids per tender, which we normalize by the size of the community (i.e., the number of firms in a community). Interestingly, Figure~\ref{Figure5}b shows that in Community 8 firms participated tend to participate in tenders with a number of participating firms that matches almost exactly the size of the community. While, in some cases---Communities 14 and 19---the numbers are several times larger. Noteworthy to mention that this analysis is biased by the size of the communities, so the expectation would be to see a smoothly increasing relationship, with the largest community achieving the smallest value, and in the limiting case of a community with a single firm we would obtain the maximum. However, it is clear that in some cases--Communities 8, 14, and 19-- there are clear deviations.

\section*{\label{conclusions}Conclusions}
In this manuscript, we explore the potential of mining a large data set of public tenders collected from the activity of firms to compete for procurement contracts issued by the municipalities of the State of Ceará (Brazil). By matching firms with similar bidding patterns, we have inferred a firm-firm network comprising a total of $1,141$ nodes and $10,630$ edges.

We show that we are able to identify communities of firms with similar bidding patterns. The network exhibits a high modular structure partitioned in $22$ communities. These communities cluster firms that have a similar scope in procurement activity both in the nature of the contracts they celebrate and in the regional reach of their activities. Moreover, we look at two diversity indicators---regional diversity and procurement contract nature diversity---as a sign of the potential of certain communities to develop leverage over the procurement process. In other words, in affecting the expected efficiency of the market. Finally, we look at the sizes of the tenders, first by looking at the abundance of single bidders in communities, and secondly by looking at the average number of bidders in each tender. Overall we have identified on a particular community (Community 8) that combines several undesirable properties. Community 8 involves a group of firms that offers Food services in the region of Jaguarabe. They have an unusually low number of single bids; the average number of participating firms per tender matches the number of firms in the community, and they exhibit a high specialization and agglomeration in their activities.

Finally, it is important to highlight some shortcomings in our analysis and future working directions. The lack of pre-labeled data on past cases of corruption largely limited our ability to make any causal link between the network structure, its motifs, and the location of firms in the network with irregular procurement behavior. In that sense, our results are merely exploratory and show the potential of combining network science methods with descriptive statistics to highlight relevant groups of firms according to their activity pattern in a data-scarce environment. Future works should look at the evolution of the network, that is, if a larger temporal window is available, to capture the evolution and segregation of communities of interest but also of their parametric path in terms of the diversity of their activities.

\section*{Acknowledgments}
    MSL was financially supported by \textit{Tribunal de Contas do Estado do Ceará} through a Phd scholarship. BD, FLP, and FB acknowledge the financial support provided by FCT Portugal under the project UIDB/04152/2020 - Centro de Investigação em Gestão de Informação (MagIC). The authors are thankful to \textit{Tribunal de Contas do Estado do Ceará} for sharing the data for this study. The authors are thankful to Cristian Candia for the useful discussions and insights. The findings, interpretations, and conclusions expressed by the authors in this work do not necessarily reflect the views of the Tribunal de Contas of Ceará.

\bibliography{References}
\end{document}